\documentclass[floats,floatfix,showpacs,preprintnumbers,amssymb,prd,twocolumn,superscriptaddress,nofootinbib,nolongbibliography,reprint]{revtex4-1}

\usepackage{amssymb,amsmath,verbatim,mathtools,needspace,enumitem,etoolbox,graphicx,physics,microtype,afterpage,xspace,tabularx,lmodern,multirow}
\usepackage{gensymb}
\usepackage[normalem]{ulem}
\usepackage[dvipsnames, usenames]{xcolor}
\usepackage{xr-hyper}
\definecolor{linkcolor}{rgb}{0.0,0.3,0.5}
\usepackage[unicode, colorlinks=true, linkcolor=linkcolor, citecolor=linkcolor, filecolor=linkcolor, urlcolor=linkcolor, linktocpage, breaklinks]{hyperref}
\usepackage[all]{hypcap}
\usepackage[T1]{fontenc}
\usepackage[utf8]{inputenc}
\usepackage{nicematrix}
\usepackage[export]{adjustbox}
\usepackage[usenames,dvipsnames]{xcolor}
\hypersetup{colorlinks=true,citecolor=romared,linkcolor=romared,urlcolor=romared}

\definecolor{romared}{RGB}{142,0,28}

\newcommand{\be}{\begin{equation}}
\newcommand{\ee}{\end{equation}}

\def\be{\begin{equation}}
\def\ee{\end{equation}}
\newcommand{\beq}{\begin{eqnarray}}
\newcommand{\eeq}{\end{eqnarray}}

\usepackage{makecell}
\usepackage{soul}

\newcolumntype{Y}{>{\centering\arraybackslash}X}

\definecolor{romared}{RGB}{142,0,28}

\makeatletter
\newcommand*{\addFileDependency}[1]{% argument=file name and extension
  \typeout{(#1)}
  \@addtofilelist{#1}
  \IfFileExists{#1}{}{\typeout{No file #1.}}
}
\makeatother

\allowdisplaybreaks

\begin{document}
\title{An exact solution describing a scalar counterpart to the Schwarzschild-Melvin Universe}

\author{Vitor Cardoso}
\affiliation{CENTRA, Departamento de F\'{\i}sica, Instituto Superior T\'ecnico -- IST, Universidade de Lisboa -- UL, Avenida Rovisco Pais 1, 1049-001 Lisboa, Portugal}
\affiliation{Niels Bohr International Academy, Niels Bohr Institute, Blegdamsvej 17, 2100 Copenhagen, Denmark}
\author{Jos\'e Nat\'ario}
\affiliation{CAMGSD, Departamento de Matem\'atica, Instituto Superior T\'ecnico, Universidade de Lisboa, Portugal}

\date{\today}

\begin{abstract}
The Schwarzschild-Melvin spacetime is an exact solution of the Einstein electrovacuum equations describing a black hole immersed in a magnetic field which is asymptotically aligned with the $z-$axis. It plays an important role in our understanding of the interplay between geometry and matter, and is often used as a proxy for astrophysical environments. Here, we construct the scalar counterpart to the Schwarzschild-Melvin spacetime: a non-asymptotically flat black hole geometry with an everywhere regular scalar field whose gradient is asymptotically aligned with the $z-$axis.
\end{abstract}

\maketitle

%\tableofcontents

%%%%%%%%%%%%%%%%%%%%%%%%%%%%%%%%%%%%%%%%%%%%
%%%%%%%%%%%%%%%%%%%%%%%%%%%%%%%%%%%%%%%%%%%%
%\section{Introduction} 
%%%%%%%%%%%%%%%%%%%%%%%%%%%%%%%%%%%%%%%%%%%%

%%%%%%%%%%%%%%%%%%%%%%%%%%%%%%%%%
\noindent {\bf \em Introduction.}
%%%%%%%%%%%%%%%%%%%%%%%%%%%%%%%%%
The Einstein equations describe with impressive depth and accuracy the gravitational interaction. The diversity of phenomena that General Relativity incorporates, from black holes to gravitational waves, is in large part due to its complex nonlinear nature. The search for exact solutions of the field equations is thus an important enterprise, allowing a deeper look into the nonlinearities, and providing explicit solutions against which to check numerical codes or test a variety of issues, such as linear and nonlinear stability.

Perhaps the most relevant -- or at least, remarkable -- nontrivial solution of Einstein equations is the Schwarzschild solution. It is remarkable in that, despite its simplicity, it describes a vacuum spacetime with such a rich content, that of a black hole geometry. 

However, the universe is filled with matter, notably electromagnetic fields. In 1964, Melvin (re-)discovered an exact solution of the Einstein electrovacuum equations (originally obtained by Bonnor~\cite{Bonnor:1954}), consisting of a simple cylindrically symmetric magnetic field aligned with the $z-$axis~\cite{Melvin:1963qx,Melvin:1965zza}, and a geometry which is not asymptotically flat:
\begin{align}
& ds^2 = D^2 \left(-dt^2 + d\rho^2 + dz^2\right) + D^{-2} \rho^2 d\varphi^2\,,\\
& \mathcal{A} = \frac{B\rho^2}{2D} d\varphi \,, \\
& D = 1 + \frac14 B^2 \rho^2 \,,
\end{align}
where $B$ is a constant and $\mathcal{A}$ is the electromagnetic $4$-potential for the Faraday tensor
\be
\mathcal{F} = d\mathcal{A} = B D^{-2} \rho \, d\rho \wedge d\varphi.
\ee
The intensity of the magnetic field (as measured by the static observers) is then $BD^{-2}$; it is maximal along the $z-$axis and decays as $\rho^{-4}$ as we move away from this axis.

The extension of this solution to include a black hole was worked out by Ernst~\cite{1976JMP....17...54E}, who obtained the line element and electromagnetic potential for the so-called Schwarzschild-Melvin solution:
\begin{align}
& ds^2=D^2\left(-fdt^2+\frac{dr^2}{f}+r^2d\theta^2\right)
+ \frac{r^2 \sin^2\theta}{D^2} d\varphi^2\,, \label{Melvin} \\
& \mathcal{A} = \frac{Br^2\sin^2\theta}{2D}d\varphi \,, \\
& D = 1 + \frac14 B^2 r^2 \sin^2\theta \,, \\
& f = 1 - \frac{2M}{r} \,.
\end{align}
This electrovacuum solution describes a black hole immersed in a magnetic field which is asymptotically aligned with the $z-$axis. It allows us to understand the motion of charged particles in a simple setup, or to study the nonlinear stability of simple magnetized spacetimes, an issue that -- despite its relevance -- is still open~\cite{PhysRev.139.B244,Brito:2014nja}.

The purpose of this note is to show that the Schwarzschild-Melvin solution admits a simple scalar counterpart, a scalar black hole universe where the gradient of the scalar field is asymptotically aligned with the $z-$axis. This universe is of interest on its own, but might be especially appealing in the context of dark matter models with a scalar or axionic degree of freedom. In addition, it might be amenable to a nonlinear stability analysis.

%%%%%%%%%%%%%%%%%%%%%%%%%%%%%%%%%
\noindent {\bf \em  The solution.}
%%%%%%%%%%%%%%%%%%%%%%%%%%%%%%%%%
We consider a minimally coupled real scalar field, described by the action
\be
S=\int d^4x\sqrt{-g}\left(\frac{R}{16\pi}-\frac{1}{2}g^{\mu\nu}\Phi_{,\mu}\Phi_{,\nu}\right)\,,
\ee
which leads to the equations of motion
\beq
R_{\mu\nu}&=&8\pi\Phi_{,\mu}\Phi_{,\nu}\,,\\
\Box\Phi&=&0\,.
\eeq
Following Buchdahl~\cite{Buchdahl:1959nk}, we look for axisymmetric solutions of these equations of the form
\begin{align}
& ds^2 = - e^{2\beta\psi} dt^2 \! + \! e^{-2\beta\psi} \left[ e^{2\gamma}\left(d\rho^2\!+\!dz^2\right) \! + \! \rho^2 d\varphi^2 \right] , \label{Weylmetric}\\
& \Phi = 2\lambda\psi \,, \label{Phipsi}
\end{align}
where the functions $\psi(\rho,z)$ and $\gamma(\rho,z)$ satisfy
\begin{align}
&\frac1{\rho}(\rho\psi_{,\rho})_{,\rho}+\psi_{,zz}=0\,,\label{Laplace}\\
&\gamma_{,\rho}=\rho\left[\left(\psi_{,\rho}\right)^2-\left(\psi_{,z}\right)^2\right]\,,\label{gamma1}\\
&\gamma_{,z}=2\rho\,\psi_{,\rho}\psi_{,z}\label{gamma2}\,,
\end{align}
and the constants $\beta$ and $\lambda$ are related by $\beta^2=1-16\pi\lambda^2$. To obtain a scalar field gradient along the $z-$axis we set
\be
\Phi=2\lambda Kz=2\lambda\psi\,,
\ee
which solves Eqs.~\eqref{Phipsi}-\eqref{Laplace}; to make the metric~\eqref{Weylmetric} independent of $z$, we are then forced to set $\beta=0$, that is, $16\pi\lambda^2=1$. Eqs.~\eqref{gamma1}-\eqref{gamma2} are now easily solved to yield $2\gamma=-K^2\rho^2$, and so we obtain the solution
\beq
ds^2&=& -dt^2 + e^{-K^2\rho^2} \left(d\rho^2 + dz^2\right) + \rho^2 d\varphi^2 \,,\\
\Phi&=&\frac{Kz}{\sqrt{4\pi}}\,.
\eeq
This is the scalar counterpart to the Melvin magnetic universe. It is ultra-static and, like the Melvin universe, it is not asymptotically flat, with spatial infinity at a finite distance along the radial direction. Unlike the Melvin universe, however, it is not globally hyperbolic, since it is possible for a light ray to reach spatial infinity in finite coordinate time. The intensity of the gradient of the scalar field is $K e^{K^2\rho^2/2}/\sqrt{4\pi}$; it is minimal along the $z-$axis and increases exponentially as we move away from this axis. Similarly to the Melvin universe \cite{PhysRevD.22.2089}, it is of Petrov type D.

To include a black hole in this scalar universe, we follow Ernst's lead and write the metric in spherical coordinates while introducing the Schwarzschild factor $f=1-2M/r$ appropriately. After some experimenting, it turns out that an exact solution is produced if we choose
\begin{align}
& ds^2 = - f dt^2 + F \left(\frac{dr^2}{f}+r^2d\theta^2\right)+r^2\sin^2\theta\,d\varphi^2 \,, \\
& \Phi = \frac{K(r-M)\cos\theta}{\sqrt{4\pi}} \,, \\
&
F = e^{-K^2r^2f\sin^2\theta} \,, \\
& f = 1 - \frac{2M}{r} \,.
\end{align}

This solution describes a scalar field whose gradient is aligned with the $z-$axis at large distances. For very small gradient $K$, we recover a linear scalar field on the background of a non-spinning Schwarzschild black hole~\cite{Berti:2013gfa}. For $K=0$ the solution reduces to the Schwarzschild geometry, whereas for $M=0$ we recover the scalar counterpart to the Melvin universe. It clearly differs from known exact solutions where scalar fields are introduced in the Schwarzschild-Melvin solution \cite{Dowker:1993bt,Agop:2005np} since, for example, such solutions cannot support scalar fields when the magnetic field is zero. Similarly to the Schwarzschild-Melvin solution \cite{PhysRevD.69.064034}, it is of Petrov type I.

%%%%%%%%%%%%%%%%%%%%%%%%%%%%%%%%%
\noindent {\bf \em Acknowledgments.} 
%%%%%%%%%%%%%%%%%%%%%%%%%%%%%%%%%
We thank Adolfo Cisterna, Roberto Emparan, Carlos Herdeiro, Roman Konoplya and David Pereniguez for very useful comments. We acknowledge support by VILLUM Foundation (grant no. VIL37766) and the DNRF Chair program (grant no.~DNRF162) by the Danish National Research Foundation.
V.C.\ is a Villum Investigator and a DNRF Chair.  
V.C. acknowledges financial support provided under the European Union’s H2020 ERC Advanced Grant “Black holes: gravitational engines of discovery” grant agreement no.~Gravitas–101052587. 
Views and opinions expressed are however those of the author only and do not necessarily reflect those of the European Union or the European Research Council. Neither the European Union nor the granting authority can be held responsible for them.
This project has received funding from the European Union's Horizon 2020 research and innovation programme under the Marie Sklodowska-Curie grant agreement no.~101007855 and no~101131233.
J.N. was partially supported by Funda\c{c}\~ao para a Ci\^encia e Tecnologia (Portugal) through CAMGSD, IST-ID
(projects UIDB/04459/2020 and UIDP/04459/2020) and project 2024.04456.CERN, and also
by the H2020-MSCA-2022-SE project EinsteinWaves, grant agreement
no.~101131233.
%
%\bibliography{References}

\begin{thebibliography}{12}%
\makeatletter
\providecommand \@ifxundefined [1]{%
 \@ifx{#1\undefined}
}%
\providecommand \@ifnum [1]{%
 \ifnum #1\expandafter \@firstoftwo
 \else \expandafter \@secondoftwo
 \fi
}%
\providecommand \@ifx [1]{%
 \ifx #1\expandafter \@firstoftwo
 \else \expandafter \@secondoftwo
 \fi
}%
\providecommand \natexlab [1]{#1}%
\providecommand \enquote  [1]{``#1''}%
\providecommand \bibnamefont  [1]{#1}%
\providecommand \bibfnamefont [1]{#1}%
\providecommand \citenamefont [1]{#1}%
\providecommand \href@noop [0]{\@secondoftwo}%
\providecommand \href [0]{\begingroup \@sanitize@url \@href}%
\providecommand \@href[1]{\@@startlink{#1}\@@href}%
\providecommand \@@href[1]{\endgroup#1\@@endlink}%
\providecommand \@sanitize@url [0]{\catcode `\\12\catcode `\$12\catcode
  `\&12\catcode `\#12\catcode `\^12\catcode `\_12\catcode `\%12\relax}%
\providecommand \@@startlink[1]{}%
\providecommand \@@endlink[0]{}%
\providecommand \url  [0]{\begingroup\@sanitize@url \@url }%
\providecommand \@url [1]{\endgroup\@href {#1}{\urlprefix }}%
\providecommand \urlprefix  [0]{URL }%
\providecommand \Eprint [0]{\href }%
\providecommand \doibase [0]{http://dx.doi.org/}%
\providecommand \selectlanguage [0]{\@gobble}%
\providecommand \bibinfo  [0]{\@secondoftwo}%
\providecommand \bibfield  [0]{\@secondoftwo}%
\providecommand \translation [1]{[#1]}%
\providecommand \BibitemOpen [0]{}%
\providecommand \bibitemStop [0]{}%
\providecommand \bibitemNoStop [0]{.\EOS\space}%
\providecommand \EOS [0]{\spacefactor3000\relax}%
\providecommand \BibitemShut  [1]{\csname bibitem#1\endcsname}%
\let\auto@bib@innerbib\@empty
%</preamble>
\bibitem [{\citenamefont {{Bonnor}}(1954)}]{Bonnor:1954}%
  \BibitemOpen
  \bibfield  {author} {\bibinfo {author} {\bibfnamefont {W.~B.}\ \bibnamefont
  {{Bonnor}}},\ }\href {\doibase 10.1088/0370-1298/67/3/305} {\bibfield
  {journal} {\bibinfo  {journal} {Proceedings of the Physical Society A}\
  }\textbf {\bibinfo {volume} {67}},\ \bibinfo {pages} {225} (\bibinfo {year}
  {1954})}\BibitemShut {NoStop}%
\bibitem [{\citenamefont {Melvin}(1964)}]{Melvin:1963qx}%
  \BibitemOpen
  \bibfield  {author} {\bibinfo {author} {\bibfnamefont {M.~A.}\ \bibnamefont
  {Melvin}},\ }\href {\doibase 10.1016/0031-9163(64)90801-7} {\bibfield
  {journal} {\bibinfo  {journal} {Phys. Lett.}\ }\textbf {\bibinfo {volume}
  {8}},\ \bibinfo {pages} {65} (\bibinfo {year} {1964})}\BibitemShut {NoStop}%
\bibitem [{\citenamefont {Melvin}(1965)}]{Melvin:1965zza}%
  \BibitemOpen
  \bibfield  {author} {\bibinfo {author} {\bibfnamefont {M.~A.}\ \bibnamefont
  {Melvin}},\ }\href {\doibase 10.1103/PhysRev.139.B225} {\bibfield  {journal}
  {\bibinfo  {journal} {Phys. Rev.}\ }\textbf {\bibinfo {volume} {139}},\
  \bibinfo {pages} {B225} (\bibinfo {year} {1965})}\BibitemShut {NoStop}%
\bibitem [{\citenamefont {{Ernst}}(1976)}]{1976JMP....17...54E}%
  \BibitemOpen
  \bibfield  {author} {\bibinfo {author} {\bibfnamefont {F.~J.}\ \bibnamefont
  {{Ernst}}},\ }\href {\doibase 10.1063/1.522781} {\bibfield  {journal}
  {\bibinfo  {journal} {Journal of Mathematical Physics}\ }\textbf {\bibinfo
  {volume} {17}},\ \bibinfo {pages} {54} (\bibinfo {year} {1976})}\BibitemShut
  {NoStop}%
\bibitem [{\citenamefont {Thorne}(1965)}]{PhysRev.139.B244}%
  \BibitemOpen
  \bibfield  {author} {\bibinfo {author} {\bibfnamefont {K.~S.}\ \bibnamefont
  {Thorne}},\ }\href {\doibase 10.1103/PhysRev.139.B244} {\bibfield  {journal}
  {\bibinfo  {journal} {Phys. Rev.}\ }\textbf {\bibinfo {volume} {139}},\
  \bibinfo {pages} {B244} (\bibinfo {year} {1965})}\BibitemShut {NoStop}%
\bibitem [{\citenamefont {Brito}\ \emph {et~al.}(2014)\citenamefont {Brito},
  \citenamefont {Cardoso},\ and\ \citenamefont {Pani}}]{Brito:2014nja}%
  \BibitemOpen
  \bibfield  {author} {\bibinfo {author} {\bibfnamefont {R.}~\bibnamefont
  {Brito}}, \bibinfo {author} {\bibfnamefont {V.}~\bibnamefont {Cardoso}}, \
  and\ \bibinfo {author} {\bibfnamefont {P.}~\bibnamefont {Pani}},\ }\href
  {\doibase 10.1103/PhysRevD.89.104045} {\bibfield  {journal} {\bibinfo
  {journal} {Phys. Rev. D}\ }\textbf {\bibinfo {volume} {89}},\ \bibinfo
  {pages} {104045} (\bibinfo {year} {2014})},\ \Eprint
  {http://arxiv.org/abs/1405.2098} {arXiv:1405.2098 [gr-qc]} \BibitemShut
  {NoStop}%
\bibitem [{\citenamefont {Buchdahl}(1959)}]{Buchdahl:1959nk}%
  \BibitemOpen
  \bibfield  {author} {\bibinfo {author} {\bibfnamefont {H.~A.}\ \bibnamefont
  {Buchdahl}},\ }\href {\doibase 10.1103/PhysRev.115.1325} {\bibfield
  {journal} {\bibinfo  {journal} {Phys. Rev.}\ }\textbf {\bibinfo {volume}
  {115}},\ \bibinfo {pages} {1325} (\bibinfo {year} {1959})}\BibitemShut
  {NoStop}%
\bibitem [{\citenamefont {Wild}(1980)}]{PhysRevD.22.2089}%
  \BibitemOpen
  \bibfield  {author} {\bibinfo {author} {\bibfnamefont {W.~J.}\ \bibnamefont
  {Wild}},\ }\href {\doibase 10.1103/PhysRevD.22.2089} {\bibfield  {journal}
  {\bibinfo  {journal} {Phys. Rev. D}\ }\textbf {\bibinfo {volume} {22}},\
  \bibinfo {pages} {2089} (\bibinfo {year} {1980})}\BibitemShut {NoStop}%
\bibitem [{\citenamefont {Berti}\ \emph {et~al.}(2013)\citenamefont {Berti},
  \citenamefont {Cardoso}, \citenamefont {Gualtieri}, \citenamefont
  {Horbatsch},\ and\ \citenamefont {Sperhake}}]{Berti:2013gfa}%
  \BibitemOpen
  \bibfield  {author} {\bibinfo {author} {\bibfnamefont {E.}~\bibnamefont
  {Berti}}, \bibinfo {author} {\bibfnamefont {V.}~\bibnamefont {Cardoso}},
  \bibinfo {author} {\bibfnamefont {L.}~\bibnamefont {Gualtieri}}, \bibinfo
  {author} {\bibfnamefont {M.}~\bibnamefont {Horbatsch}}, \ and\ \bibinfo
  {author} {\bibfnamefont {U.}~\bibnamefont {Sperhake}},\ }\href {\doibase
  10.1103/PhysRevD.87.124020} {\bibfield  {journal} {\bibinfo  {journal} {Phys.
  Rev. D}\ }\textbf {\bibinfo {volume} {87}},\ \bibinfo {pages} {124020}
  (\bibinfo {year} {2013})},\ \Eprint {http://arxiv.org/abs/1304.2836}
  {arXiv:1304.2836 [gr-qc]} \BibitemShut {NoStop}%
\bibitem [{\citenamefont {Dowker}\ \emph {et~al.}(1994)\citenamefont {Dowker},
  \citenamefont {Gauntlett}, \citenamefont {Kastor},\ and\ \citenamefont
  {Traschen}}]{Dowker:1993bt}%
  \BibitemOpen
  \bibfield  {author} {\bibinfo {author} {\bibfnamefont {F.}~\bibnamefont
  {Dowker}}, \bibinfo {author} {\bibfnamefont {J.~P.}\ \bibnamefont
  {Gauntlett}}, \bibinfo {author} {\bibfnamefont {D.~A.}\ \bibnamefont
  {Kastor}}, \ and\ \bibinfo {author} {\bibfnamefont {J.~H.}\ \bibnamefont
  {Traschen}},\ }\href {\doibase 10.1103/PhysRevD.49.2909} {\bibfield
  {journal} {\bibinfo  {journal} {Phys. Rev. D}\ }\textbf {\bibinfo {volume}
  {49}},\ \bibinfo {pages} {2909} (\bibinfo {year} {1994})}\BibitemShut
  {NoStop}%
\bibitem [{\citenamefont {Agop}\ \emph {et~al.}(2005)\citenamefont {Agop},
  \citenamefont {Radu},\ and\ \citenamefont {Slagter}}]{Agop:2005np}%
  \BibitemOpen
  \bibfield  {author} {\bibinfo {author} {\bibfnamefont {M.}~\bibnamefont
  {Agop}}, \bibinfo {author} {\bibfnamefont {E.}~\bibnamefont {Radu}}, \ and\
  \bibinfo {author} {\bibfnamefont {R.}~\bibnamefont {Slagter}},\ }\href
  {\doibase 10.1142/S0217732305017317} {\bibfield  {journal} {\bibinfo
  {journal} {Mod. Phys. Lett. A}\ }\textbf {\bibinfo {volume} {20}},\ \bibinfo
  {pages} {1077} (\bibinfo {year} {2005})}\BibitemShut {NoStop}%
\bibitem [{\citenamefont {Ortaggio}(2004)}]{PhysRevD.69.064034}%
  \BibitemOpen
  \bibfield  {author} {\bibinfo {author} {\bibfnamefont {M.}~\bibnamefont
  {Ortaggio}},\ }\href {\doibase 10.1103/PhysRevD.69.064034} {\bibfield
  {journal} {\bibinfo  {journal} {Phys. Rev. D}\ }\textbf {\bibinfo {volume}
  {69}},\ \bibinfo {pages} {064034} (\bibinfo {year} {2004})}\BibitemShut
  {NoStop}%
\end{thebibliography}
%
%\clearpage
%\renewcommand{\thesubsection}{{S.\arabic{subsection}}}
%\setcounter{section}{0}

%%%%%%%%%%%%%%%%%%%%%%%%
%
%%%%%%%%%%%%%%%%%%%%%%%%

\end{document}